# Measuring Fermi velocities with ARPES in narrow band systems. The case of layered cobaltates.

V. Brouet[1], A. Nicolaou[1,2], M. Zacchigna[3], A. Taleb-Ibrahimi[2], P. Le Fèvre[2] and F. Bertran[2]

[1] *Lab. Physique des Solides, Université Paris-Sud, UMR8502, Bât 510, 91405 Orsay, France*
[2] *Synchrotron SOLEIL, L'Orme des Merisiers, Saint-Aubin-BP48, 91192 Gif-sur-Yvette, France*
[3] *CNR-INFM, Laboratorio Nazionale TASC c/o Area Science Park - s.s. 14 Km. 163.5, I-34012 Basovizza (TS), Italy*

ARPES is *a priori* a technique of choice to measure the Fermi velocities $v_F$ in metals. In correlated systems, it is interesting to compare this experimental value to that obtained in band structure calculations, as deviations are usually taken as a good indicator of the presence of strong electronic correlations. Nevertheless, it is not always straightforward to extract $v_F$ from ARPES spectra. We study here the case of layered cobaltates, an interesting family of correlated metals. We compare the results obtained by standard methods, namely the fit of spectra at constant momentum $k$ (energy distribution curve, EDC) or constant binding energy $\omega$ (momentum distribution curve, MDC). We find that the difference of $v_F$ between the two methods can be as large as a factor 2. The reliability of the 2 methods is intimately linked to the degree of $k$- and $\omega$-dependence of the electronic self-energy. As the $k$-dependence is usually much smaller than the $\omega$ dependence for a correlated system, the MDC analysis is generally expected to give more reliable results. However, we review here several examples within cobaltates, where the MDC analysis apparently leads to unphysical results, while the EDC analysis appears coherent. We attribute the difference between the EDC and MDC analysis to a strong variation of the photoemission intensity with the momentum $k$. This distorts the MDC lineshapes but does not affect the EDC ones. Simulations including a $k$ dependence of the intensity allow to reproduce the difference between MDC and EDC analysis very well. This momentum dependence could be of extrinsic or intrinsic. We argue that the latter is the most likely and actually contains valuable information on the nature of the correlations that would be interesting to extract further.

## 1. Introduction

In many correlated systems, a small quasiparticle (QP) peak appears near the Fermi level ($E_F$), which is quickly lost at higher binding energies, as it broadens and/or loses weight. This is for example the case in weakly doped cuprates [1], manganites [2-4] or Na [5-6] and misfit [7-8] cobaltates. In Fig. 1, such a situation is presented in the misfit cobaltate $[Bi_2Ba_2O_4][CoO_2]_2$ [7]. The peak near $E_F$ corresponds to the crossing of a hole-like $a_{1g}$ band [8,9]. Its dispersion is clear up to 50meV binding energies. It is very important to find the right way to extract the characteristics of this QP peak (Fermi velocity and broadening) in order to characterize the properties of the system. In this paper, we will show that this analysis is sometimes tricky and should be done carefully. Especially, it has already been noted that fitting lines of the image corresponding to constant momentum k (energy distribution curve EDC) or constant binding energy $\omega$ (momentum distribution curve MDC) may give quite different results. In Fig. 1b, the dispersion is extracted by the two types of analysis (see details below) and the Fermi velocities $v_F$ differ from 0.28eV.Å in EDC analysis to 0.38eV.Å MDC analysis. In some cases, the difference can be even larger creating serious problems to estimate important quantities such as the $v_F$ renormalization, which is a fundamental quantity to evaluate the strength of correlations. The origin of this discrepancy has not been clearly elucidated so far and it is therefore not known which method should be trusted.

In this paper, we show that this difference can be assigned to a *k*-dependence of the peak intensity, which distorts the MDC lineshapes. For a given *k*-dependence of the intensity, we show that the effect is stronger in systems with low Fermi velocities and/or large peak width. We will show several examples, where changes in $v_F$ extracted by MDC analysis are clearly unphysical. We will discuss possible origins for the *k*-dependence of the intensity, either extrinsic or intrinsic. Extrinsic origins would be related to *k*-dependence of the photoemission matrix elements. However, the dependence observed here seems too strong to be attributed to this effect alone. Intrinsic origins would be related to a *k*-dependence of the self-energy, i.e. non-local correlations. To our knowledge, there are not many examples where such correlations could be isolated in ARPES spectra, so that it would be interesting to find ways indicating their potential importance. The case we document here – a strong deviation between $v_F$ obtained in EDC or MDC analysis - may contribute to this task.

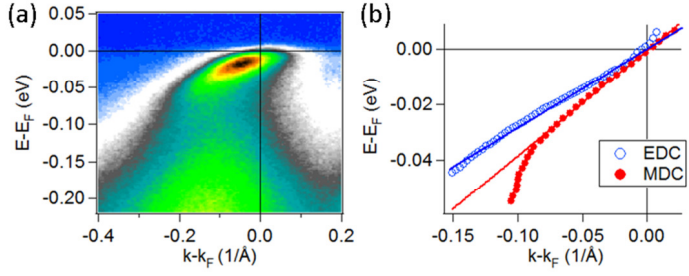

Figure. 1 :
(a) ARPES energy-momentum intensity plot in the misfit cobaltate BiBaCo acquired at 20K with a photon energy of 55eV and linear polarization in the plane of light incidence. The direction of the reciprocal space was ΓM and $k_F = 0.57 Å^{-1}$. (b) Dispersion extracted by EDC (open blue symbols) or MDC (closed red symbols) analysis. Solid lines are linear fits of the extracted dispersions near $E_F$, yielding Fermi velocities of 0.28 and 0.38eV.Å in the MDC or EDC cases, respectively.

## 2. Experimental

We present results from $Na_{0.8}CoO_2$ and 3 misfit cobaltates, abbreviated below as BiBaCo, CaCoO and Bi(Sr,Pb)Co. In Na cobaltates, the triangular $CoO_2$ planes are stacked with Na layers. In misfit cobaltates, the same $CoO_2$ planes are stacked with rock-salt layers that may be incommensurate. Their composition is A-O / Bi-O / Bi-O /A-O for Bi misfit cobaltates with A=Ba for BiBaCo and A=$Sr_{0.72}Pb_{0.28}$ for Bi(Sr,Pb)Co. For CaCoO, there are 3 intercalated layers Ca-O/Co-O/Ca-O. More information on their properties can be found in ref. 7, 10 or 11. For Na and misfit cobaltates, the electronic structure is essentially that of the $CoO_2$ plane, to which the intercalated planes transfer electrons. The triply degenerate $t_{2g}$ band of Co is filled with 5 to 6 electrons. Only one band is found to cross the Fermi level, forming a hexagonal hole-like Fermi Surface centered at Γ [5−8].

Single crystals were prepared by floating zone method for Na cobaltates and solid flux reaction for misfit cobaltates. Experiments were carried out with synchrotron light, at the CASSIOPEE beamline of SOLEIL, the SIS beamline of the Swiss Light Source and the APE beamline of ELETTRA. In each case, the data were acquired with a SCIENTA analyser, the energy resolution was typically 10meV and the angular resolution 0.3°.

## 3. Model

The ARPES intensity is proportional to the single particle spectral function A(k,ω) through :

$$I(k, \omega) = \left|M_{A,h\nu}(k)\right|^2 A(k, \omega) f(\omega) \qquad (1)$$

where $f(\omega)$ is the Fermi function and $M_{A,h\nu}(k)$ a matrix element describing the transition probability for the photoelectron from its initial to final state after absorbing a photon [12,13]. This matrix element depends on quantities related to the photon beam (polarization A and photon energy hν), but also on the electron momentum k (for example through its orientation compared to A). On the other hand, it is not expected to depend much on the binding energy ω, which is usually safely negligible compared to hν [14]. As it is quite complicated to calculate such matrix elements, they are often ignored. However, they sometimes dramatically modulate photoemission intensities, as in some cuprates [15]. One question we will address in this paper is whether the k-dependence of these matrix element effects can be safely neglected in all cases.

Typically, one is interested to extract the spectral function from I(k,ω), because it is directly related to the electron self-energy Σ(k,ω) that contains information about the interactions among electrons.

$$A(k,\omega) = \frac{-1}{\pi} \frac{\Sigma''(k,\omega)}{[\omega - \varepsilon_k - \Sigma'(k,\omega)]^2 + [\Sigma''(k,\omega)]^2} \qquad (2)$$

It can be expressed into a coherent and incoherent part, by developing the function around its pole $\omega_p(k) = \varepsilon_k + \Sigma'(k, \omega_p)$ [12,13].

$$A(k,\omega) = Z_k \frac{Z_k \Sigma''(k,\omega_p)/\pi}{[\omega - \omega_p(k)]^2 + [Z_k \Sigma''(k,\omega_p)]^2} + (1 - Z_k) A_{inc}(k,\omega) \qquad (3)$$

The first term describes a Quasiparticle (QP) of weight $Z_k = \left|1 - \frac{\partial \Sigma'}{\partial \omega}(k, \omega_p)\right|^{-1}$, energy $\varepsilon_k + \Sigma'(k, \omega_p)$ and lifetime $\left(Z_k \Sigma''(k, \omega_p)\right)^{-1}$. The second term is a priori unknown, it can range from a satellite to a continuous background. As correlations increase, one expects the QP weight to decrease, the linewidth to increase and the dispersion to be more and more strongly renormalized. All these evolutions are contained in $\Sigma(k,\omega)$. The spectra are typically strongly $\omega$-dependent; in a Fermi liquid, a lifetime scaling as $\omega^2$ is expected. On the other hand, it is usually a good approximation to neglect the k-dependence of the self-energy. It will only become sizable, if there are strong non-local correlations, for example inter-site antiferromagnetic or charge couplings.

Consequently, one can expect a spectrum at constant $\omega$ (i.e. the MDC) to be much simpler than a spectrum at constant k (i.e. the EDC). In fact, Eq. 3 shows that it is a simple lorentzian, if the self-energy exhibits no k-dependence and the dispersion is modeled as linear $\omega = v_F (k-k_F)$. An additional advantage of the MDC analysis is that it is independent of the Fermi function and has a more simple background. All this was first recognized by T. Valla *et al.* [16] and has led to an intensive use of the MDC analysis, preferred over EDC analysis [17,18]. This indeed allowed reaching a much better accuracy in extracting dispersions and linewidth, revealing kinks and other important effects [12]. However, fitting MDC with lorentzians is only valid if the variation of intensity as a function of k can be neglected. This is also well known in the community, but there is no easy way to check whether this assumption is valid. Eq. 1 and Eq. 3 recall that this means neglecting both extrinsic variations through matrix element effects and intrinsic variations through $Z_k$ and $\Sigma''(k,\omega)$. To go beyond the MDC/EDC analysis, many authors have recently proposed two-dimensional analysis [19], for example relating directly $\Sigma'$ and $\Sigma''$ by Kramers-Kronig transformation [20]. To our knowledge, there is however no clear cut case identified so far, where a k-dependence of the self-energy could be unambiguously deduced from ARPES spectra, and shown to invalidate the MDC analysis. In the following, we adopt simple analysis methods (namely, EDC and MDC fits) to compare spectra exhibiting different deviations between MDC and EDC $v_F$ values, in order to evaluate their physical origin.

In Fig. 2, typical MDC and EDC spectra extracted from the image of Fig. 1 are displayed. We already note that the MDC are obviously not perfect lorentzians, suggesting some caution should be taken in applying the MDC analysis. Nevertheless, a lorentzian fit allows to extract a dispersion $k(\omega)$ and a width $\Gamma(\omega)$, which are reported as solid points in Fig. 3a and 3b. To fit the EDC, we chose in Fig. 2b a lorentzian on top of a linear background, divided by the Fermi function. The Fermi level was measured on a reference gold sample and the resolution was fixed to 12meV. The results of this analysis (QP dispersion, broadening and weight) are presented in Fig. 3. We do not expect this fit to be accurate, as it assumes that the linewidth can be defined at each k-value, i.e. that its variation with $\omega$ can be neglected on the spectral width. This is a rough approximation for a correlated Fermi liquid, where it is expected to change as $\omega^2$, especially near $E_F$.

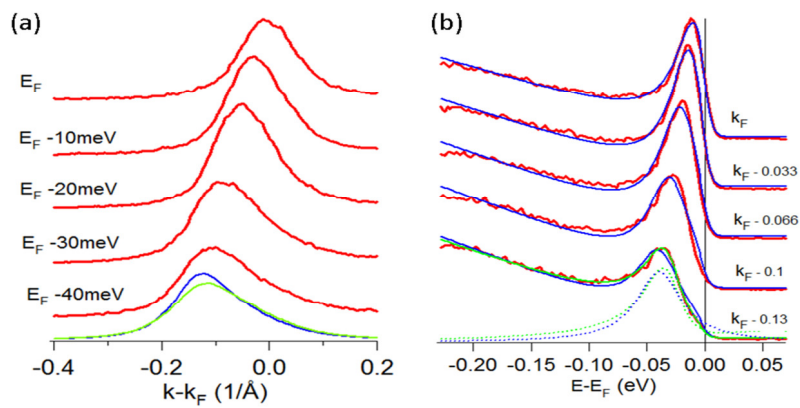

Figure 2 : Thick red lines show typical MDC (left) and EDC (right) spectra extracted from the image of Fig. 1 at the indicated energies or k values. In (a), the MDC at $E_F$-40meV corresponding to the two simulations discussed in the text are shown. In (b), fits used to extract the QP properties are shown as blue lines. They consist of a lorentzian on top of a linear background. For the bottom spectra, another fit is shown in green, where the linewidth of the QP peak is a $\omega^2$ function of the energy. At bottom, the QP contribution in the two fits are shown.

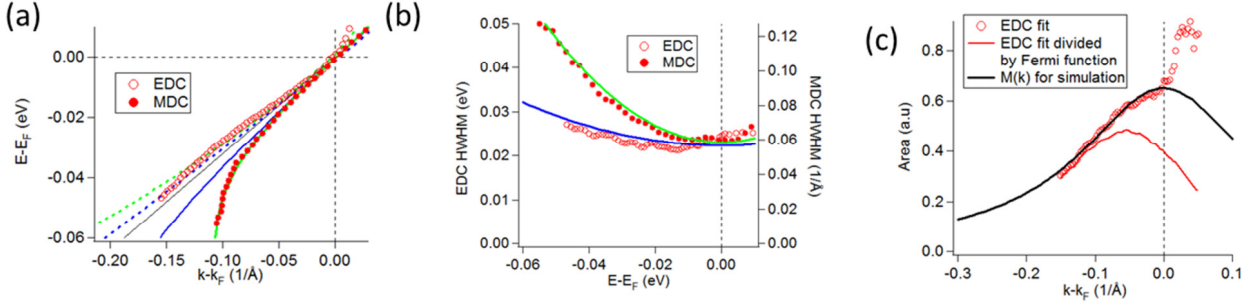

Figure 3 : (a) Dispersions obtained for EDC and MDC fits. The thin black line is the dispersion used in the simulation. The thick dotted (resp. solid) lines are the fit obtained by EDC (resp. MDC) analysis of the simulated images. (b) Linewidth obtained in EDC and MDC fits. Blue and green lines are the models used for the simulation (see text). (c) Intensity as a function of k obtained through the EDC fit (open symbols). The blue line shows the area divided by the Fermi function. The thick black line is the model for I(k) used in the simulation (see text).

However, Fig. 2b shows that it gives a reasonable starting point to describe the lineshapes. At the bottom of the image, we show an alternative fit (green line), where the linewidth is taken as $\Gamma=A+B\omega^2$. This of course improves the fit, but this does not significantly change the area of the QP (bottom dotted lines), which is the main quantity we are interested in here. Note that, whereas, in the Fermi liquid theory, the lineshape should strongly distort from a lorentzian as one approaches $E_F$, the residual linewidth due to impurity scattering and the convolution with experimental resolution will efficiently mask this distortion. This explains why assuming lorentzian EDC works reasonably well.

The results for the dispersion in Fig. 3a evidences the difference already reported in Fig. 1b, which is the main point we want to explore in this paper. Note also that there is a strong "kink" for the MDC dispersion around -0.04eV, which is not seen in the EDC. As for the linewidths (Fig. 3b), they are similar near $E_F$, after normalizing the EDC width (in eV) by the slope of the dispersion (the ratio of the left and right scale in Fig. 3b is about the MDC $v_F$). However, their values differ at higher binding energies. On one hand, the MDC fit is certainly wrong for E>0.05eV, where it does not fit the QP anymore, but tails of the structures at higher binding energies (see Fig. 1 and [8]). On the other hand, the EDC fit is quite dependent on the function chosen for fitting (a simple lorentzian here) and should not be very reliable if the width is strongly ω dependent. As we will see, this difference is not crucial to simulate the difference between EDC and MDC and we will not try to determine the linewidth more accurately in this paper.

In Fig. 3c, we report the variation of the area observed as a function of k. For this, we have to rely on the EDC analysis, which should include variation of $M_{A,h\nu}(k)$ and $Z_k$ (the MDC intensity I(ω) has a completely different meaning and should just correspond to the density of states multiplied by the Fermi function). This variation is quite steep, being divided by a factor 2 over 0.15Å$^{-1}$. We also show the area divided by the Fermi function to evidence that the intensity variation is already quite clear, before part of the intensity is cut off by the Fermi function. We will return later to possible meanings of this variation. For the time being, we just assume such a variation I(k) and see whether it helps to simulate the difference between EDC and MDC fittings. Qualitatively, the higher intensity near the Fermi level will tend to distort the MDC towards the occupied side of the band. This is exactly the behavior observed in Fig. 2a. The MDC are quite lorentzian near $E_F$ but with a strong tail towards $k_F$ at higher binding energies. We assume the following distribution of spectral intensity.

$$I(k,\omega) = I(k) \frac{\Gamma(\omega)}{v_F^2(k-k(\omega))^2 + \Gamma(\omega)^2} \qquad (4)$$

As for the area, we model I(k) by the black line in Fig. 3c, i.e. a lorentzian centered at $k_F$ and of width 0.18Å$^{-1}$. There is no particular reason to choose such a function, it is just convenient. The dispersion is taken as linear with $v_F$=0.28eV.Å, as from the EDC fit. As the ω dependence of Γ is not very well known, we consider two functions (see Fig. 3b) : $\Gamma(\omega)$ =0.06+6ω$^2$ (blue line) and $\Gamma(\omega)$ =0.06+15ω$^2$ (green line). We have found that the result of the simulation are not very sensitive to the shape of Γ(ω), but mainly to its value near $E_F$, which are quite similar in the two analysis.

With these parameters, we built images I(k,ω) that we fit using standard EDC and MDC analysis. We find that this allows reproducing very well the different $v_F$ observed experimentally. Fig. 3a shows that the slope of the MDC near $E_F$ does not depend much on the form of Γ(ω) and that they are indeed steeper than those found for EDC. These latter values correspond to the input values, despite the different forms chosen for EDC in the model and in the fit. Note also, that the MDC dispersion is not linear but displays a "kink" at higher binding energies, which is created by the way the linewidth changes and depends consequently sensitively on the form chosen for Γ(ω). Typical spectra obtained in the simulation are shown at the bottom of Fig. 2a and indeed reproduce the spectra well.

We conclude that including a k-dependence of the QP intensity allows to reproduce the difference between EDC and MDC. Especially, there is no need to include a background due to higher binding energy structures to reproduce this difference. This latter effect is also present, and it is probably important in the kink region, but it forms a rather broad background, which does not seem to modify the behaviors near $E_F$. At this point, we have not determined which method is the most reliable. One could favor the MDC analysis because the lorentzian fit is more justified and attribute the I(k) variation found in the EDC analysis to an improper fitting form. Alternatively, if the I(k) variation is real, one should favor the EDC fit, because the EDC lineshape is not distorted by this effect. In the following, we examine real experimental situations to find out which analysis seems able to capture the physics behind the data.

## 4. Application to spectra with different width

In Fig. 4a, we compare dispersions extracted in BiBaCo in three different situations. Two spectra (triangles and circles) were obtained at 5K on the same sample and in the same experimental conditions (12meV resolution), but at different sample positions. As shown in Fig. 4b, they appear to have slightly different width. The circles (BiBaCo1) correspond to the case discussed before and is a typical width encountered in BiBaCo. The triangles (BiBaCo2) corresponds to a particularly narrow spectrum, as we have sometimes encountered. While it is quite understandable that the width may somewhat vary across the sample (the number of impurities may for example be different), it seems wrong that dispersions also change. As shown in Fig. 4a or 4c, $v_F$ is almost the same in EDC analysis (open symbols) but increases steeply in the MDC analysis. Next, we consider a spectrum taken at higher temperature (200K), where the spectrum is significantly broader. Here again, while the broadening is expected from the physics, one would not expect a strong change in the dispersion of this compound. If it does, one would rather expect a narrowing of the band at higher temperatures, i.e. a smaller $v_F$, corresponding to a loss of coherence at high temperature and a stronger renormalization. On the contrary, the MDC values increase significantly. On the other hand, the EDC value is quite constant, within fitting accuracy.

The model presented before explains very well these differences. Indeed, the MDC will pick more and more intensity from the high intensity side in samples with broader spectra, and therefore distort more and more strongly. To simulate this effect, we calculated $v_F$ obtained through MDC analysis for a constant dispersion ($v_F=0.28$eV·Å) but different linewidth at $E_F$. As $v_F$ is not very sensitive to the energy dependence of the linewidth, we just assumed a constant linewidth. We obtained the red line in Fig. 4c, which is indeed quite consistent with the variation of the MDC. We observe that the deviation is negligible for peak width $\Gamma_{MDC}<0.04$Å$^{-1}$. This is about one fourth of the distance $\Gamma_k$ needed to divide the intensity by 2 in the model (0.15 Å$^{-1}$ in Fig. 3c). We find that this ratio remains constant if we use different k-variations or different peak width, so that a criterion for the validity of MDC analysis is $\Gamma_{MDC}<\Gamma_k/4$ or $\Gamma_{EDC}<\Gamma_k*v_F/4$. Unfortunately, it is not straightforward to define $\Gamma_k$, so that the practical use of this criterion is limited.

In this example, we would conclude that the changes in $v_F$ indicated by the MDC analysis are meaningless. Although less precise and more difficult to model, the EDC analysis should be preferred in these situations, because it captures the right physics.

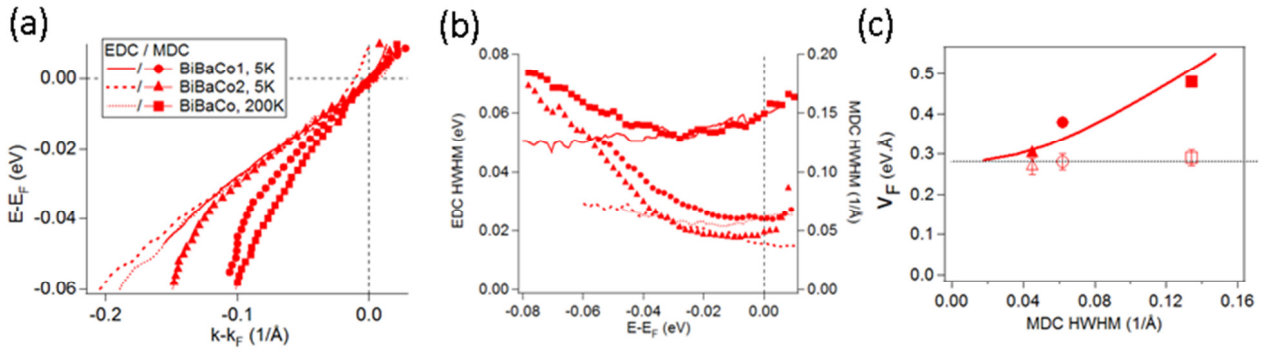

Figure 4 : Dispersion (a) and linewidth (b) obtained in one BiBaCo sample at 5K and two different sample positions (called 1 and 2 in caption), as well as at 200K. The experimental conditions are the same than in Fig. 1. Symbols refer to MDC fits and lines to EDC fits. (c) $v_F$ obtained for EDC (open symbols) and MDC (filled symbols) for the previous cases as a function of the MDC linewidth at $E_F$. The thick line indicates $v_F$ obtained in a simulation considering $v_F=0.28$eV·Å, the intensity variation of Fig. 3a and a constant linewidth.

## 5. Application to spectra in different compounds

We now aim at comparing the dispersions in different cobaltates. Fig. 5a shows a collection of EDC spectra taken at k values just below $k_F$ in different compounds. They exhibit quite different QP shapes, which might be due to a broadening and/or a loss of QP intensity. In blue, we show a spectrum in $Na_{0.8}CoO_2$, which is quite typical of the ones found in the literature [5,6,21], although much narrower spectra have also been reported [22]. The fact that spectra with quite different widths have been reported suggest that surface quality plays a role in the linewidth. Generally, cleaving is more difficult in $Na_xCoO_2$ than in Bi misfit cobaltates, where samples cleave easily between two Bi-O layers. Consequently, surfaces are generally of poorer quality and certainly age faster. This may at least play a role in the difference of spectra between BiBaCo and $Na_xCoO_2$. In the other misfits, the QP also appears weaker than in BiBaCo. The surface quality may also be different (especially, BiBaCo is the only commensurate compound), but this evolution seems related to an increase in doping, the QP peak being completely suppressed at higher dopings [7]. We will not attempt here to interpret the evolution of the QP properties, this will be reported in details elsewhere [23], but we focus on how to estimate and compare their properties, notably their dispersions.

The dispersions extracted by EDC or MDC analysis are reported in Fig. 5b. Here again, the evolution is strikingly opposite in the two cases. The MDC analysis finds a steep *increase* of $v_F$ compared to BiBaCo in the compounds with the smallest peaks, while the EDC fits finds similar or weaker dispersion [especially in Bi(Sr,Pb)Co]. Intuitively, one would rather expect that a smaller peak is a sign of stronger correlation effects, so that $v_F$ should rather decrease. This would agree with the EDC analysis and strongly disagrees with the MDC evolution. The EDC fit becomes in fact very difficult when the peak weakens, especially near $E_F$. The evolution of $v_F$ is then difficult to define precisely, and it might be more reliable to define a quantity proportional to the bandwidth by the value of the EDC fit at $k=-0.2Å^{-1}$. We obtain 70meV in BiBaCo, 60meV in $Na_{0.8}CoO_2$, 50meV in CaCoO and just 30meV in BiSrPbCo, indicating as narrowing of the bandwidth in qualitative agreement with the reduction of the peak. On the other hand, we conclude again that the MDC results give unphysical results here.

In Fig. 5c, we plot the values of Fermi velocities as a function of the MDC linewidth. This quantity is the easiest to define and corresponds to the inverse mean free path. Here, we observe that the difference in EDC/MDC $v_F$ is not entirely controlled by the linewidth (the red line is the linewidth dependence determined for BiBaCo in section 4), but depends on the compounds. Indeed, the MDC linewidth are quite similar for $Na_{0.8}CoO_2$, CaCoO and BiSrPbCo, but the EDC/MDC difference sharply increase in CaCoO and BiSrPbCo. While the value for $Na_{0.8}CoO_2$ is of the order of the one expected in BiBaCo, it is completely different in the other compounds. This means that the physics of the latter compounds is likely different, for example the intrinsic $v_F$ and/or I(k) has changed. In fact, it even seems that stronger correlations (defined by smaller peak and narrower bandwidth, like for CaCoO and BiSrPbCo) directly leads to a stronger EDC/MDC deviation, so that this deviation could even be taken as a sign of the increase of the correlations.

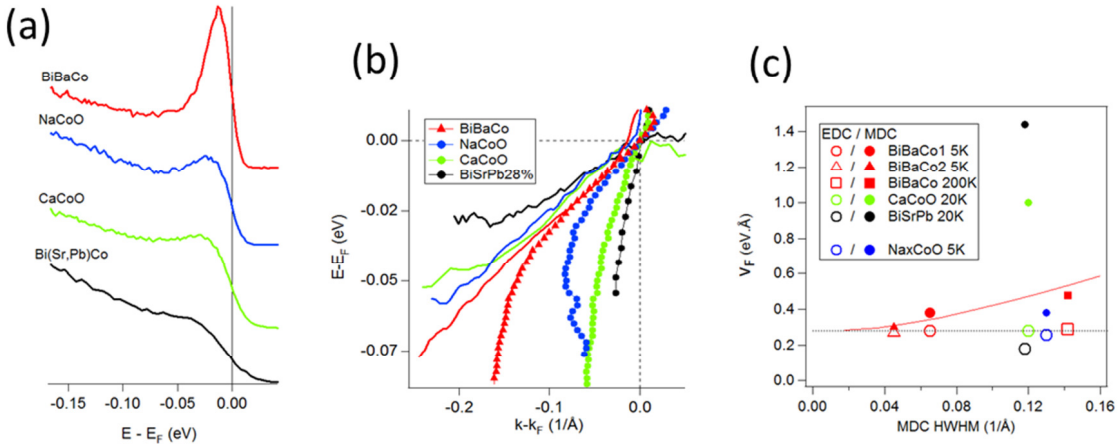

Figure 5 : (a) Typical EDC spectra just below $k_F$ (-10meV binding energy) in different cobaltates for photon energies around 100eV and linear polarization in the plane of light incidence. (b) Dispersion extracted with EDC (solid line) or MDC (symbols) analysis. (c) $v_F$ values as a function of MDC linewidth in the previous cases.

## 6. Discussion

The origin of the k-dependence of the intensity is obviously at the heart of a better understanding of these lineshapes. It may be extrinsic and associated with k-dependent matrix-element effects (see Eq. 1) or intrinsic and associated with k-dependence of the QP weight $Z_k$ (see Eq. 3). As for extrinsic effects, a strong k-dependence could be anticipated as the $a_{1g}$ orbital is quite anisotropic and therefore, the intensity should depend sensitively on how the polarization is oriented compared with $a_{1g}$. However, one would typically expect a cosine square variation [7], which should still be relatively smooth over the angular range covered by the QP peak dispersion (at 50eV, 0.2Å$^{-1}$ corresponds to only 3°). Therefore, it seems rather difficult to attribute all the dependence to such a variation. Moreover, we have found for BiBaCo the same dispersions and lineshapes for measurements between 20 and 100eV.

We then have to consider an intrinsic variation of $Z_k$. This is in fact quite likely in cobaltates. For example a spin-polaron theory has been proposed to describe the lineshape, in which Z strongly changes with k [24] and the role of non-local correlations to explain the complicated charge orders observed there has been directly emphasized [25]. When the k-dependence of the self-energy can be neglected, $v_F$ is simply renormalized by Z (one has Z=m/m*, as can be simply deduced from the equations in section 3). This, however, would only hold in a relatively narrow energy window near $E_F$ [26]. With a k-dependence, Z and the $v_F$ renormalization may become quite different. This means that, even if the renormalization of $v_F$ could be clearly defined, it would not necessarily capture the full extent of correlations. In ref. 6, we have proposed a method to identify the incoherent excitations clearly in cobaltates and deduce Z from the transfer of spectral weights. Such a method is rarely used in ARPES, because it is apparently much more difficult to control than a simple fit of the dispersion, but this example suggests that it could also be very useful and complementary.

Finally, we would like to comment on shortcomings of our model. We believe that it is useful to identify the k-dependence of the peak intensity as the origin of the deviation between EDC and MDC values in a phenomenological way. However, it is far too simple to extract the k-dependence of the self-energy (even assuming that matrix element effects can be neglected). First, if $Z_k$, which depends on the real part of the self-energy $\Sigma'(k,\omega)$ (see section 3), is k-dependent, there must be an associated k-dependence of the imaginary part $\Sigma''(k,\omega)$, directly distorting the MDC linewidth. This should be taken into account, which is not the case in our model assuming in eq. 4 that Γ does not depend on k. Second, the renormalization of the dispersion also depends on $\Sigma'(k,\omega)$, so that it is not likely to remain linear, if $Z_k$ is changing. This is the origin of the "kinks" observed when Z changes suddenly. In the case of a coupling to phonons, for example, the kink at the Debye frequency $\omega_D$ is indeed associated to a change in Z from m/m* below $\omega_D$ (m* being the effective mass renormalized by electron-phonon coupling) to 1 above [27]. In the general case, the dispersion might deviate smoothly from the linear dispersion, even if the bare dispersion itself was truly linear. All these effects should be taken consistently by assuming a general form from $\Sigma(k,\omega)$, satisfying the Kramers-Kronig relations. We feel there are too many unknown parameters, especially the bare band dispersion and the small ω window where $\Sigma'$ and $\Sigma''$ can be measured, to undertake such an analysis reliably. The k-dependence of the peak intensity is the easiest feature to extract experimentally and we believe it is a good starting point to evaluate the impact of k-dependent self-energy on the spectra.

## 7. Conclusion

We have shown that the discrepancy between EDC and MDC can be simply explained by including a k-dependence of the QP intensity. We have shown two examples where this dependence distorts the MDC linewidth so much that fitting it yields unphysical results. Despite its poorer accuracy, the EDC fitting remains more reliable in these cases. To trust the MDC analysis, one should make sure that the variation of I(k) is negligible over the typical peak width in k-space. More precisely, in our simulations, the variation of intensity should be smaller than 30% over the peak width to keep the deviation negligible. Unfortunately, it is usually not easy to determine if this condition is fulfilled. On the other hand, we suggest that a large deviation between MDC and EDC, such as the one we report in CaCoO and BiSrPb in Fig. 5c, can be taken as the sign of very significant k-dependent effect. More work will be needed to find a proper way to extract this information from the data.

Non-local correlations are typically neglected in many theoretical approaches, most notably the Hubbard model, but it seems more and more clear that they will be needed to describe certain situations. In cobaltates, a great mystery is that correlations remain strong – or even increase - away from half-filling [7]. Including non-local correlations may be a way to answer this problem [25], but there is a lack of experimental tools to support such ideas. We have underlined that there may be very different contributions to the deviation of $v_F$ in EDC/MDC analysis, but we also suggest it could signal an importance role of k-dependent effects in one compounds.

On a more practical side, these findings should be kept in mind when comparing spectra in different compounds. In Na cobaltates, for example, it would be extremely interesting to understand how correlations evolve from the low doping to the high doping regime, but the use of $v_F$ determined by MDC alone [6] appears very questionable. It is also quite important to consider such effects to analyze the variation as a function of temperature.

## 8. Acknowledgements

We thank L. Patthey, M. Shi and I. Vobornik for help during measurements on the SIS and APE beamlines, P. Lejay for the $Na_{0.8}CoO_2$ crystal, H. Muguerra and S. Hébert for the misfit cobaltates crystals, A. Georges, F. Lechermann and O. Peil for useful discussions.